# Correlated Ligand Electrons in the Transition-Metal Oxide SrRuO$_3$

Yuichi Seki[1], Yuki K. Wakabayashi[2], Takahito Takeda[1], Kohdai Inagaki[1], Shin-ichi Fujimori[3],

Yukiharu Takeda[3], Atsushi Fujimori[3,5,6], Yoshitaka Taniyasu[2],

Hideki Yamamoto[2], Yoshiharu Krockenberger[2], Masaaki Tanaka[1,4,7], and Masaki Kobayashi[1,4,*]

[1]*Department of Electrical Engineering and Information Systems, The University of Tokyo, Bunkyo, Tokyo 113-8656, Japan*

[2]*NTT Basic Research Laboratories, NTT Corporation, Atsugi, Kanagawa 243-0198, Japan*

[3]*Materials Sciences Research Center, Japan Atomic Energy Agency, Sayo-gun, Hyogo 679-5148, Japan*

[4]*Center for Spintronics Research Network, The University of Tokyo, 7-3-1 Hongo, Bunkyo-ku, Tokyo 113-8656, Japan*

[5]*Center for Quantum Technology and Department of Physics, National Tsing Hua University,*

*Hsinchu 30013, Taiwan*

[6]*Department of Physics, The University of Tokyo, Tokyo 133-0033, Japan*

[7]*Institute for Nano Quantum Information Electronics, The University of Tokyo, 4-6-1 Komaba, Meguro-ku, Tokyo 153-8505,*

*Japan*

*Corresponding author: masaki.kobayashi@ee.t.u-tokyo.ac.jp

**Abstract**

In transition-metal compounds, the transition-metal $d$ electrons play an important role in their physical properties; however, the effects of the electron correlation between the ligand $p$ electrons have not been clear yet. In this Letter, the Ru 4$d$ and O 2$p$ partial density of states (PDOS) in transition-metal oxide SrRuO$_3$ involving Weyl fermions are investigated by resonant photoemission spectroscopy. The observations demonstrate that the O 2$p$ PDOS is distorted from that predicted by first-principles calculations than the Ru 4$d$ PDOS. The results indicate that the electron correlation in the ligand orbitals will be important to understand the electronic structure of the $p$-$d$ hybridized state in strongly correlated electron systems, even with topological states.

# Introduction

Transition-metal compounds known as strongly correlated electron systems have been intensively studied due to a variety of their physical properties such as superconductivity [1], metal-insulator transition [2, 3, 4], and multiferroics [5, 6]. The electron correlation between the transition-metal $d$ electrons affects the electronic structure near the Fermi level ($E_F$) and plays an important role in their physical properties [1-4, 7, 8]. Additionally, the interplay between electronic correlations and topological states has attracted attention as correlated topological phases for engineering the topological properties [9-15]. In high-$T_c$ cuprate and copper oxides, photoemission studies have suggested that the values of electron correlation in O $2p$ are about 5–6 eV [16, 17]. Recently, in addition to the $d$ electrons, the effects of the electron correlation between the ligand $p$ electrons have been discussed theoretically and experimentally in transition-metal compounds; electron-correlation in O $2p$ orbitals varying in different families in cuprates [18], observation of the delocalized ligand Ge/Te $p$ orbitals in van der Waals magnet FeGeTe [19], spin-polarized electronic state on the ligand N $2p$ orbital in GaN:Eu [20]. However, it has not been clear yet how the ligand electron correlation affects the physical properties of transition-metal compounds.

Transition-metal oxide $SrRuO_3$ (SRO), which crystallizes in a pseudocubic perovskite structure as shown in Fig. 1(a), has been intensively investigated for more than 60 years due to its unique physical properties such as metallic transport and perpendicular magnetic anisotropy [3, 21-27]. In this study, SRO is a model system for clarifying the relationship between the electronic state of the ligand O $2p$ and its macroscopic physical properties since the recent improvement of the crystallinity in the thin films allowed us to access the intrinsic electronic structure of the coherent Ru $4d$-O $2p$ hybridized states [28-31]. Indeed, thanks to the ultra-high crystal quality, the quantum transport phenomena of Weyl fermions in ultrahigh-quality thin films have recently been observed [32-37]. In addition to the quantum transport measurements [34, 38], combinations of first-principles calculations with other experiments such as neutron scattering [39] and angle-resolved photoemission spectroscopy [40] for SRO suggest the existence of the Weyl nodes in the electronic structure near $E_F$. Figure 1(b) illustrates the crystal-field splitting and the spin configuration of the Ru 4$d$ orbitals. Because of the octahedral ($O_h$) crystal field formed by the ligand O ions, the degenerate Ru 4$d$ orbitals are split into $e_g$ and $t_{2g}$ orbitals with the low-spin configuration ($t_{2g}\uparrow^3$, $t_{2g}\downarrow^1$) [3, 41].

In this paper, we have elucidated the effect of the correlated ligand electrons by direct observations of the Ru 4$d$ and O 2$p$ partial density of states (PDOS) in 4$d$ transition-metal oxide SRO using resonant photoemission spectroscopy (RPES). Here, there have been few reports of the O $K$ RPES on transition-metal oxides [17, 42, 43] because the resonant photoemission process is principally ineffective for the nearly closed shell configuration of the ligand element, especially for O with high electronegativity. The experimental results demonstrate that the Ru 4$d$ PDOS is significantly different from the O 2$p$ PDOS in the $p$-$d$ hybridized state near $E_F$, which is against the general belief that the feature of $p$ PDOS is similar to that of $d$ PDOS due to the $p$-$d$ hybridization, as predicted by first-principles calculations [44, 45]. The electron-electron Coulomb energy between the ligand O 2$p$ electrons ($U_{pp}$) estimated experimentally is much larger than that between the Ru 4$d$ electrons ($U_{dd}$) in SRO. The resonant enhancement of the O 2$p$ PDOS at the strong Ru 4$d$-O 2$p$ XAS peak due to the ultrahigh crystal quality of the film allows us to evaluate the effect of $U_{pp}$ in a transition-metal oxide quantitatively for the first time. Based on the findings, we discuss the role of the correlated ligand O 2$p$ for the emergence of the Weyl fermions in SRO. This provides experimental evidence that the electron correlation between the $p$ electrons makes the PDOS of the metal-$d$ and ligand-$p$ orbitals significantly different near $E_F$.

**Experimental**

An ultrahigh-quality SRO film with its thickness of 60 nm was grown on a SrTiO$_3$ (STO) (001) substrate by a custom-designed molecular beam epitaxy (MBE) setup equipped with multiple e-beam evaporators of Sr and Ru. Here, the growth parameters were optimized by Bayesian optimization, which is a machine learning technique for parameter optimization [33, 34], resulting in the fabrication of an ultrahigh-quality SRO thin film with residual resistivity ratio RRR = 56. Generally, the purity and crystalline quality of metallic electronic systems are evaluated by RRR and large RRR values are required to observe quantum transport phenomena [34, 46]. The compressive strain on the SRO layer from the STO substrate was about 0.6% [32] (see Appendix I for the crystal structure). Soft X-ray absorption spectroscopy (XAS) and photoemission spectroscopy (PES) measurements were performed at the undulator beamline BL23SU of SPring-8. The monochromator energy resolution $E/\Delta E$ was about 10 000 and the beam spot size was < 200 μm in diameter. The PES measurements were conducted with an SES-2002 electron analyzer (Gammadata-Scienta Co.Ltd.) at 28 K at a base pressure of $1.7 \times 10^{-8}$ Pa. The XAS spectra were taken in the total-electron yield (TEY) mode. In the PES measurements, the overall energy resolutions including the thermal broadening were 103 and 110 meV for the incident photon energies ($h\nu$) of 526 and 529 eV, and 115 and 135 meV for 454 and 463 eV with narrower slit width, respectively. The position of $E_F$ was calibrated by measuring evaporated Au in electrical contact with the sample.

**Results and discussion**

To elucidate the Ru 4$d$ states in SRO, we have conducted the XAS and RPES at the Ru $M$ edge. Figure 2(a) shows the Ru $M_{2,3}$-edge XAS spectrum of the SRO film. The line shape of the Ru $M_{2,3}$ spectrum well coincides with previous reports on ultrahigh-quality SRO films [29, 30]. Additionally, the core-level spectrum is comparable to the electronic structure determined earlier (See Fig. 5 in Appendix). The energy positions of $M_3$ ($h\nu \sim$ 463 eV) and $M_2$ ($h\nu \sim$ 486 eV) absorption peaks correspond to the transitions from Ru $3p_{3/2}$ and Ru $3p_{1/2}$ core levels into the unoccupied Ru 4$d$ states, respectively [47].

As shown in Fig. 2(b), the Ru $M_3$ on- and off-resonance PES spectra are taken at $h\nu =$ 463 and 454 eV, respectively [indicated in Fig. 2(a)]. The spectral line shapes demonstrate an intense peak structure derived from the coherent quasiparticle states near $E_F$, which is one of the characteristic features of high-quality SRO thin films [29, 30, 48]. The intensity of the Ru 4$d$ coherent quasi-particle state is enhanced at $h\nu =$ 463 eV on resonance. As shown in the lower part of Fig. 2(b), the Ru 4$d$ PDOS has been obtained as the difference between the on- and off-resonant spectra. The sharp peak in the spectrum reflects the coherent states near the $E_F$. With the increase of RRR value, the coherent quasiparticle peak becomes more intense [48] and, therefore, the observation of the prominent coherent peak reflects the high crystallinity of the measured film. Considering the reports from several theoretical and experimental studies [29, 30, 44, 49, 50], the coherent states observed near $E_F$ probably correspond to the Ru 4$d$ $t_{2g}$–O 2$p$ anti-bonding states ($t_{2g}^a$). The existence of the coherent states in the Ru 4$d$ PDOS indicates that the Ru 4$d$ electrons predominantly contribute to the transport properties of SRO.

In addition to the Ru 4$d$ states, the O 2$p$ states in the valence-band structure near $E_F$ have been analyzed using XAS and RPES at the O $K$ edge. Figure 3(a) shows the O $K$-edge XAS spectrum around the energy region corresponding to the absorption into the unoccupied Ru 4$d$–O 2$p$ hybridized states. From the similarity of the spectral line shapes between the present and the previous XAS spectra on ultrahigh-quality SRO films [29, 30], the absorption peak at 529 eV is assigned to transitions into $t_{2g}^a$ and the structures in the energy range of 530-534.5 eV are ascribed to transitions into Ru 4$d$ $e_g$–O 2$p$ hybridized states. The intensity of the Ru 4$d$ $t_{2g}$ absorption peak is higher and sharper than that of the Ru 4$d$ $e_g$ components determined in previous studies [49, 51-55], reflecting the long lifetime of quasiparticles in the $t_{2g}^a$ orbital near $E_F$. This is consistent with the previous studies on ultrahigh-quality SRO thin films [29, 31, 33], providing us with further evidence for the high crystallinity of the measured sample. Figure 3(b) shows the O $K$ RPES spectra near $E_F$ taken with varying incident photon energies in the O $K$ XAS region. In comparison of the RPES spectra with the off-resonance spectrum ($h\nu$ = 526 eV), the resonant enhancement is maximized approximately at $h\nu$ = 529 eV. The $h\nu$ dependence of the PES intensity suggests the on-resonance energy is $h\nu$ = 529 eV. Since the resonant enhancement in PES (Super Coster-Kronig transition) requires hole states in the valence band of the objective element, this result indicates that the ligand O 2$p$ band has a finite number of holes [56, 57]. As this enhancement occurs in parallel with the electron excitation into the unoccupied states in $t_{2g}^a$ orbitals, the existence of the hole states suggests a portion of O 2$p$ orbitals becomes unoccupied and composes part of the holes in the $t_{2g}^a$ orbitals. Figure 3(c) shows the O $K$ on- and off-resonance PES spectra taken at $h\nu$ = 529 and 526 eV, respectively. On-resonance spectrum has two enhanced features, one at $E_B \geq 2.7$ eV and the other at $0 \leq E_B \leq 2.5$ eV. The former is ascribed to the excitation from the O 2$p$ non-bonding band [47, 49, 53] and a peripheral part of O ($KLL$) Auger emission [18, 43, 55, 58, 59] in which the peak position shifts with $h\nu$ (see Fig. 7(a) in Appendix). Figure 3(d) shows the O 2$p$ PDOS as the difference between the on- and off-resonance spectra. It should be noted here that the O 2$p$ PDOS in the $t_{2g}^a$ band lies between $0 \leq E_B \leq 2.5$ eV. Compared with the Ru 4$d$ PDOS shown in Fig. 2(b), the O 2p PDOS shows a relatively broader peak centered around $E_B$ = 1.5 eV with a finite but low intensity at $E_F$.

In the RPES measurements, the Ru 4d PDOS (Fig. 2(b)) and the O 2p PDOS (Fig. 3(d)) in $t_{2g}{}^a$ band near $E_F$ have been observed. Based on our experimental findings, the electronic structure of SRO realized by the Ru 4d $t_{2g}$–O 2p hybridization is schematically illustrated in Fig. 4(a). From the different spectral shapes of the Ru 4d and O 2p PDOSs, it is probable that the Ru 4d and the O 2p PDOSs unequally contribute to the $t_{2g}{}^a$ bands as indicated in Fig. 4(a): The Ru 4d PDOS has coherent states near the $E_F$ leading to the itinerant nature. In contrast, the O 2p PDOS gradually diminish toward $E_F$ and demonstrate small DOS near the $E_F$. That is pseudogap behavior observed in strongly correlated electron systems [3, 4]. The non-bonding and the bonding bands ($t_{2g}{}^b$) are also observed in the valence-band PES spectra (See Fig. S7(b) in Appendix).

Here we discuss the possible scenario for the difference in the observed Ru 4d and O 2p PDOS spectra. The pseudogap structure of the O 2p PDOS can be explained by the effect of electron correlations, as is the case for Mott-Hubbard-type insulators [2]. In detail, O 2p–derived $t_{2g}{}^a$ electrons are affected by relatively strong electron correlations compared to the Ru 4d-derived $t_{2g}{}^a$ electrons, as discussed below. Considering the p-d hybridization, the wave function $\Psi$ of the ground states for the RuO$_6$ cluster in SRO is given by:

$$\Psi = C_0|d^4\rangle + C_1|d^5\underline{L}\rangle + C_2|d^6\underline{L}^2\rangle + \cdots, (1)$$

where $\underline{L}$ denotes a hole in the O 2p ligand band and $C_i$ ($i$ = 0, 1, 2, …) is the linear combination coefficients. $|d^{n+i}\underline{L}^i\rangle$ configurations are realized by charge transfer from the O 2p to Ru 4d bands via strong hybridization. Since the charge-transfer states involve states with a finite number of holes in the O 2p band, the influence of electron correlation on such states with holes in the O 2p band is expected to be finite for the states near $E_F$, whereas the O 2$p$ electron correlation in the $|d^4\rangle$ state with fully occupied O 2$p$ band is ineffective [60]. Considering the mixture of the terms in Eq. (1) (called as configuration interaction), the O 2$p$ PDOS in the $t_{2g}{}^a$ band is caused by the cross terms with the charge-transfer states. Hence, it is expected that the angular-orbital ($p$ or $d$) dependent distributions of the PDOS originate from the presence or absence of the charge-transfer states. Furthermore, the electron correlations of the O 2$p$ likely affect the distribution of the PDOS in the $t_{2g}{}^a$ band. To verify this hypothesis, the on-site Coulomb energy of the O 2$p$ electrons $U_{pp}$ is estimated as $U_{pp}$ = 5.7 ± 0.1 eV using the Cini-Sawatzky method (See Fig. 7(c) in Appendix) and found to be significantly larger than that of Ru 4$d$ $U_{dd}$ = 1.7 ~ 3.1 eV [61-63]. Based on this relationship and the multiple configurations in $\Psi$, $t_{2g}{}^a$ band can be schematically described as shown in Fig. 4(b). In contrast to coherent Ru 4$d$ PDOS, the larger $U_{pp}$ causes the pseudogap structure in the O 2$p$ PDOS (cross terms with the charge-transfer states in $|\Psi|^2$). It follows from these arguments that the difference between the Ru 4$d$ and O 2$p$ PDOS in $t_{2g}{}^a$ band comes from the different electron correlations depending on the elemental orbital characters.

    Based on our observation, we can further discuss the physical properties of SRO. The pseudo-gap behavior of the O 2$p$ PDOS near $E_F$ indicates that the O 2$p$ state hardly contributes to the transport properties of SRO and the high itineracy of the quasiparticle around $E_F$ predominantly comes from the Ru 4$d$ state [29]. As for the quantum transport phenomena of SRO, the emergence of the Weyl fermion in the $p$-$d$ hybridized state is related to the balance between the Coulomb interaction and the band width [9, 10]. Generally, topological states with linear band dispersion are weakly correlated because of the relatively large band widths compared with the Coulomb interactions. Considering the strong $U_{pp}$ of the O 2$p$ PDOS, it is possible that the hybridization with the O 2$p$ negatively acts for the emergence of the Weyl fermion. However, in the energy

range near $E_F$ that the Weyl nodes exist, the O 2$p$ PDOS with the pseudogap behavior less contributes to the transport property than the Ru 4$d$ PDOS, Thus, the emergence of the Weyl nodes mainly originates from the inherent characteristics of the Ru 4$d$ PDOS even in the $p$-$d$ hybridized state and the hybridization with O 2$p$ may contribute to extend the band width of the hybridized state.

Additionally, the presence of the charge-transfer states $|d^{n+i}\underline{L}^i\rangle$ can also explain the non-trivial finite magnetic moment on the ligand O ions observed by X-ray magnetic circular dichroism [29, 30]. Since Ru in the SRO (001) has spin-polarized 4$d$ electrons [21, 41], the ligand electrons with minority spins tend to be involved in the charge-transfer states from the fully-occupied O 2$p$ orbitals. It is probable that this spin-selective charge transfer from the O 2$p$ band induces spin-polarization in the O 2$p$ band, which explains the non-trivial magnetic moment on the ligand O ions in SRO.

Traditionally, it has been considered that electron correlation between $d$ electrons $U_{dd}$ plays an important role in the physical properties of strongly correlated electron systems. Indeed, in previous theoretical studies for transition-metal compounds, only $U_{dd}$ is considered for electron correlation, leading to basically similar distributions of PDOS irrespective of the orbital characteristics ($p$ or $d$) in the $p$-$d$ hybridized states. [33, 36, 37, 43, 50, 62, 63]. As discussed above, the present observations demonstrate that the spectral line shape of the $d$ PDOS is dramatically different from that of the $p$ PDOS in the $p$-$d$ hybridized states because of the large Coulomb energy of the ligand $p$ states $U_{pp}$ compared to that of the transition-metal metal $d$ states $U_{dd}$. This consequence raises a question about the conventional description of transition metal $d$ – ligand $p$ orbital hybridization, that is, the uniform mixture of $d$- and $p$-electrons in the hybridized states, in particular, in 4$d$ and 5$d$ transition-metal oxides, where $U_{dd}$ is relatively small. This consideration will be generally applied to understanding the electronic structures of ligand elements in correlated *p-d* hybridized systems [18, 19, 64, 65]. The results suggest that further intensive theoretical and experimental studies on the electronic structure caused by ligand hybridization will extend our understanding of physical backgrounds effective in the fascinating properties of the correlated ligand materials.

**Conclusion**

In conclusion, we have performed XAS and RPES measurements on an ultra-high quality SRO thin film to elucidate the ligand electron correlation. The observations demonstrate that the Ru 4*d* PDOS show a different spectral line shape from the O 2*p* PDOS even in their hybridized states in the vicinity of $E_F$: The Ru 4*d* PDOS establish an intense coherent peak across $E_F$ and, therefore, predominantly contributes to the itinerant electronic behavior, whereas the O 2*p* PDOS turns out to show a pseudogap-like spectral line shape in the vicinity of $E_F$. This result is different from theoretical calculations showing the relatively uniform distributions of the PDOSs in the hybridized state, where the *p* electrons are assumed to have negligible electron correlation compared to the *d* orbitals. Based on our experimental findings, the dramatically different *d* and *p* PDOSs in the *p-d* hybridized state are explained by the charge transfer from the Ru 4*d* $t_{2g}$ orbital to the O 2*p* orbital and the electron correlation of the ligand O 2*p* orbitals which is stronger than that of the Ru 4*d* orbitals. Actually, the value of $U_{pp}$ estimated by the Cini-Sawatzky method is double of $U_{dd}$ predicted from theoretical calculations. It is likely that the non-trivial finite magnetic moment on the ligand $O^{2-}$ ions originates from the spin-selective charge transfer from the O 2*p* band to the spin-polarized Ru 4*d* band. As for the quantum transport property of SRO, the O 2*p* PDOS with pseudogap behavior less contribute to the transport property due to the relatively strong $U_{pp}$ and the emergence of the Weyl fermion probably originates from the inherent characteristics of the Ru 4$d$ PDOS in the $p$-$d$ hybridized state.

The present results suggest a general perspective that the transition-metal $d$ and ligand $p$ PDOSs in the $p$-$d$ hybridized states possibly depend on their orbital characteristics in transition-metal compounds. This provides experimental evidence that the electron correlation between the $p$ electrons makes PDOS of the metal-$d$ and ligand-$p$ orbitals significantly different near $E_F$. Based on the present findings, the electron correlation in the ligand orbital, especially for ligand elements with high electron negativity such as O and N, will play an important role in understanding the electronic structure of the $p$-$d$ hybridized states in strongly correlated electron systems.

# Appendix

## I. Crystal structure of SrRuO$_3$/SrTiO$_3$ thin films

A previous study of structural phase transition for SrRuO$_3$ (SRO)/SrTiO$_3$ (STO) substrate has revealed that the strained-Cubic phase is changed to the orthorhombic phase below 280 ºC [66]. Additionally, it is known that bulk SRO remains in the orthorhombic phase from 300 K to 12 K, according to low-temperature X-ray diffraction (XRD) measurements [67]. Therefore, it is expected that epitaxial SRO on STO also retains the orthorhombic phase from 300 K down to 28 K, across the ferromagnetic transition temperature around 150 K. The epitaxial SRO film is grown with out-of-plane [110] direction of the orthorhombic phase with [001] and [-110] directions of the orthorhombic phase aligned in-plane on the STO (001) substrate [68]. We note that the [110], [001], and [-110] directions of the orthorhombic phase correspond to the pseudocubic [001], [100], and [010] directions, respectively.

## II. Core-level X-ray photoemission spectroscopy spectra

Figure 5 shows the Sr 3p and Ru 3d core-level PES spectrum. The peaks at 285.5 eV and 278.3 eV are ascribed to C 1s and Sr $3p_{1/2}$ [51, 54, 69]. The C 1s component is likely derived from the air contamination. The Ru $3d_{5/2}$ components are observed from 280.1 to 284.2 eV. The sharp peak around 281.1 eV is formed due to the screening of the hole state in Ru 3d by itinerant electrons (called as well-screened peak) [54, 70]. The broad peak around 282.5 eV is formed due to poorly screening. Compared with the previous reports [51, 54], our PES spectrum demonstrates a more prominent and sharper well-screened peak relative to the poorly-screened peak. As well as the previous study on an ultrahigh-quality SRO thin film [29], the high-itinerant electrons in SRO are likely to cause the prominent peak, which is consistent with the high crystallinity of our sample.

Figure 6 shows the wide scan spectrum of an SRO/STO thin film. The stoichiometry estimated by the spectral areas is Sr:Ru:O ~ 1:1:5. This result suggests that the layers near the surface are O-rich. The O 1s spectrum likely includes both the intrinsic O atoms in SRO and extrinsic contaminations ($H_2O$, CO, *etc*). Actually, the finite C 1s peak due to the surface contamination was observed. It is probable that the presence of the surface contamination hardly contributes to the main conclusion of this study because the surface contamination forms bonding states well below the Fermi level ($E_F$) and then they have negligible density of states in the vicinity of $E_F$.

## III. Decomposition of valence band spectrum and details of the estimation of $U_{pp}$

Figure 6(a) shows a valence-band spectral mapping around the O K absorption edge (~ 529 eV) and O K XAS spectrum corresponding to that energy region. One can see the enhancement of the PES intensity occurs from $hv$ ~ 529 eV, where the X-ray absorption into the unoccupied Ru 4d – O 2p anti-bonding state ($t_a$) in the XAS spectrum. For $hv \geq 529$ eV, an intense KLL auger emission (See Fig. 6(b)) appears and its peak position shifts to higher binding energy ($E_B$) as $hv$ increases (shown with a dotted line in Fig. 6(a)). The significant intensity of the Auger emission is derived from the larger cross-section of the KLL Auger process than that of direct photoemission from the valence band.

Figure 6(c) shows the average of the valence-band spectra observed with $454 \leq hv \leq 470$ eV. The averaged valence band spectrum mainly consists of the quasiparticle states ($0 \leq E_B \leq 2.3$ eV) derived from Ru 4d and the convolutions of multiple electronic structures ($2.3 \leq E_B \leq 10$ eV) mainly derived from the O 2p bands [49, 50, 71]. For the estimation of electron correlation in the O 2p orbital (written in the following paragraph), the PES spectrum in the range of $2.3 \leq E_B \leq 10$ eV is numerically decomposed into three Gaussians. Two of them located around $E_B = 6$ eV are attributed to the Ru 4d - O 2p bonding states ($t_b$), and the peak around $E_B = 3.7$ eV is supposed to be the O 2p non-bonding states [49].

Here we estimate the electron correlation effect in the O 2p orbital $U_{pp}$ based on the Cini-Sawatzky method [18, 43, 72, 73]. In Fig. 6(d), we compare the Auger emission spectrum taken at $hv = 529$ eV with the numerical self-convolution of the O 2p states including the non-bonding and bonding states. In this method, the self-convolution of the one-hole PES spectrum can be regarded as the two-hole spectrum and the energetic difference between the Auger emission and the self-convolution spectra gives the electron correlation $U$. While the non-bonding states are dominantly occupied by the O 2p PDOS [50, 71], the ratio of the Ru 4d density of states to the O 2p bonding state is about 40% according to a first-principles theoretical calculation [71]. Additionally, the photoionization cross section of Ru 4d is 30 times larger than that of O 2p at $hv = 1250$ eV [74]. Then, the total intensity of the bonding state includes the contribution of the Ru 4d state as 92%.

By extracting the Ru 4$d$ contribution from the bonding state, we regard the estimated $U$ from the self-convoluted O 2$p$ PDOS. The energy separation between the main peaks of the Auger and the self-convoluted spectra corresponds to $U_{pp}$ and then the value of $U_{pp}$ is estimated as 5.7 ± 0.1 eV, as shown in Fig. 6. This estimated $U_{pp}$ is close to the other estimations in the different kinds of perovskite oxides [18, 43, 75], high-$T_c$ cuprate, and copper oxides of about 5–6 eV [16, 17] in previous RPES studies. The value of $U_{pp}$ is larger than the electron correlation in the Ru 4$d$ orbital ($U_{dd}$) in the ruthenates ranging from 1.7 to 3.1 eV estimated from experiments and calculations [45, 61, 63].

## Acknowledgment

The authors are grateful to H. Katayama-Yoshida for informative discussion. This work was supported by Grants-in-Aid for Scientific Research Grants-in-Aid for Scientific Research (19K21961, 20H05650, 22H04948, and 22K03535), CREST program (JPMJCR1777) of Japan Science and Technology Agency. This work was partially supported by the Spintronics Research Network of Japan (Spin-RNJ). This work was carried out under the Shared Use Program of Japan Atomic Energy Agency (JAEA) Facilities (Proposal No. 2022A-E23) supported by JAEA Advanced Characterization Nanotechnology Platform as a program of "Nanotechnology Platform" of the Ministry of Education, Culture, Sports, Science and Technology (MEXT) (Proposals No. JPMXP1222AE0031). The experiment at SPring-8 was accepted by the Japan Synchrotron Radiation Research Institute (JASRI) Proposal Review Committee (Proposal No. 2020A3841).

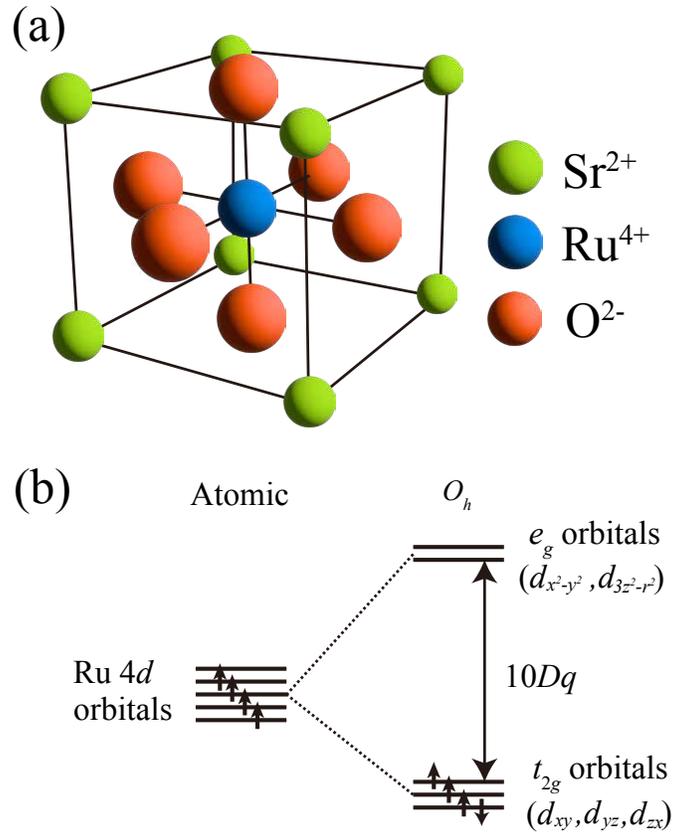

Fig.1 Schematic view of the perovskite oxide SrRuO$_3$ (SRO). (a) Perovskite crystal structure of SRO. (b) Electron and spin configurations in the degenerate Ru $4d$ orbitals in an isolated Ru atom and their separation due to the octahedral crystal field ($O_h$) in SRO. Here, $10Dq$ is the crystal-filed splitting. A black arrow on the energy levels represents an electron with a spin.

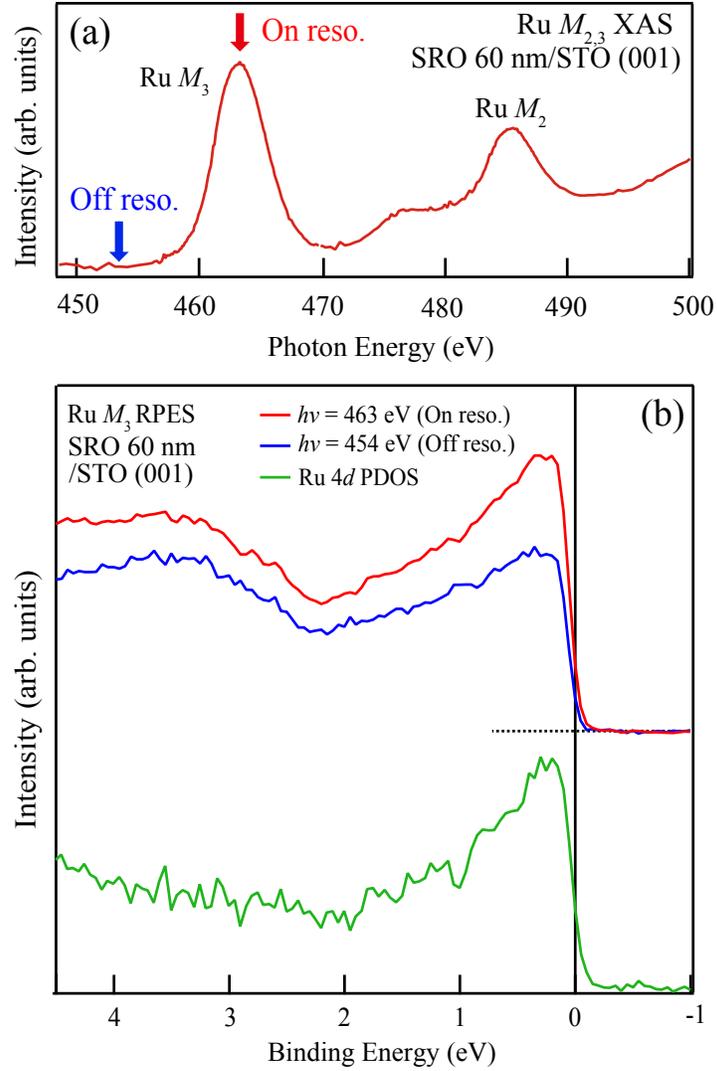

Fig.2 Ru $M_{2,3}$ XAS and $M_3$ PES spectra of an ultrahigh-quality SrRuO$_3$/SrTiO$_3$ thin film. (a) Ru $M_{2,3}$ XAS spectrum. Red and black arrows show the photon energies at which on- and off-resonance spectra were taken. (b) Ru $M_3$ RPES spectra. The on- and off-resonant spectra have been measured at $h\nu$ = 463 and 454 eV, respectively. The Ru 4$d$ PDOS is obtained as the difference between the on- and off-resonance spectra.

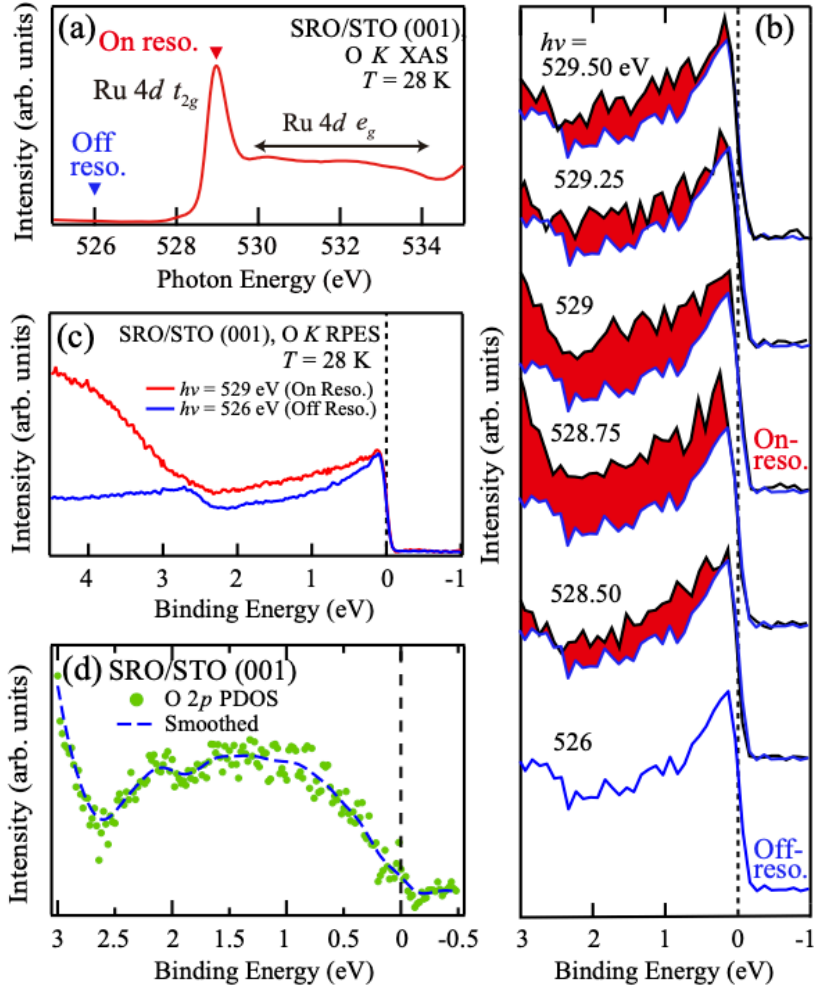

Fig.3 O $K$ XAS and RPES spectra of an ultrahigh-quality SrRuO$_3$/SrTiO$_3$ thin film. (a) O $K$-edge XAS spectrum. (b) Series of RPES spectra of the valence band. The spectra are taken at photon energies indicated as downward triangles in (a) ($hv$ = 526 ~ 529.50 eV. The red areas are positive differences relative to the off-resonance spectrum. (c) Comparison between the on- ($hv$ = 529 eV) and off-resonance ($hv$ = 526 eV) valence-band PES spectra. (d) O $2p$ PDOS obtained as the difference between the on- and off-resonant valence-band spectra. The smoothed line for the data of the O $2p$ PDOS is also shown.

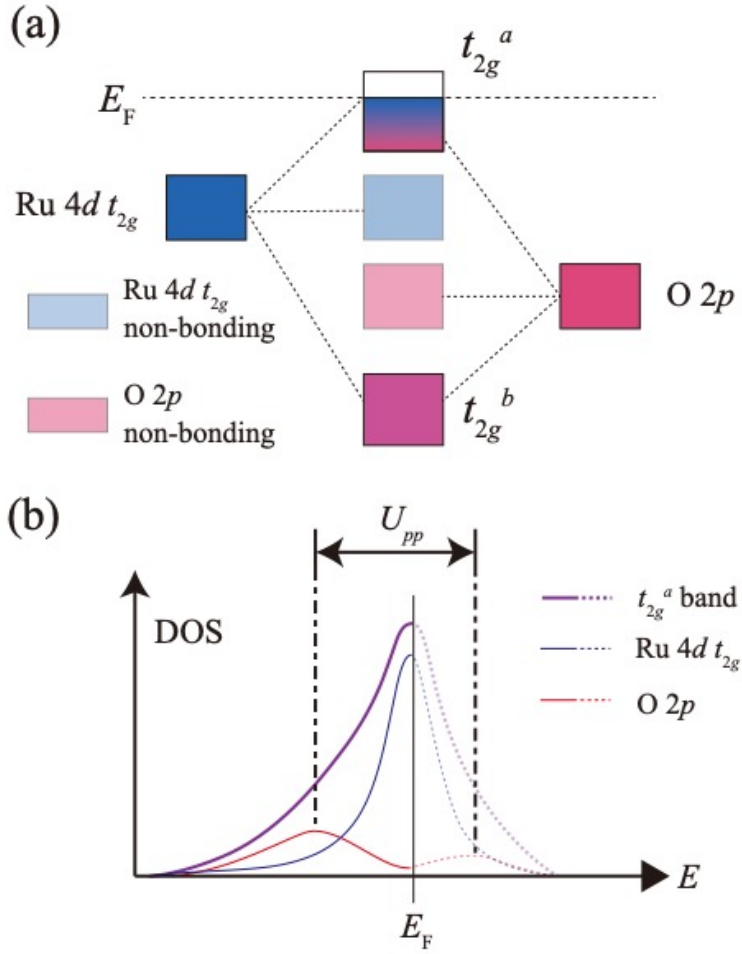

Fig.4 Band diagram of SrRuO$_3$. (a) Schematic energy diagram of the valence-band structure through the Ru 4$d$–O 2$p$ hybridization in SRO. "$t_{2g}^a$" and "$t_{2g}^b$" indicate the anti-bonding and bonding bands, respectively. The gradation in the $t_{2g}^a$ band describes the transition of the dominant components of the Ru 4$d$ and O 2$p$ electrons, i.e., the orbital-dependent distributions of Ru 4$d$ and O 2$p$ PDOSs. (b) Schematic picture of the Ru 4$d$ and O 2$p$ PDOS in $t_{2g}^a$ band near $E_F$. The DOS is drawn with solid line below $E_F$ and dotted line above $E_F$. The $U_{pp}$ indicates the on-site Coulomb energy for electrons in the O 2$p$ band.

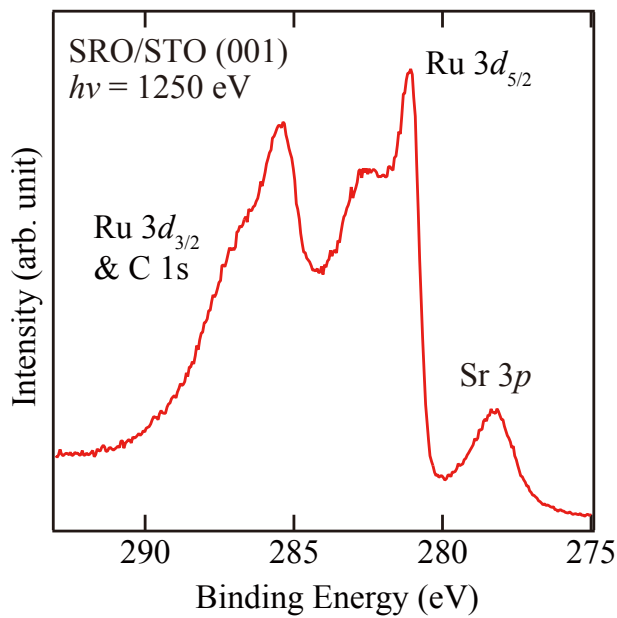

Fig. 5. Ru 3*d* and Sr 3*p* core level PES spectrum of the SRO thin film taken with $h\nu = 1250$ eV.

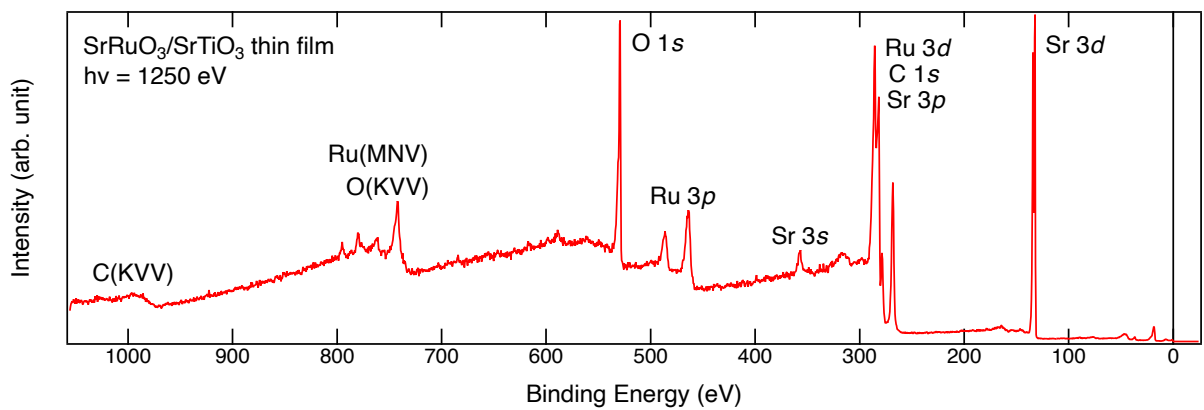

Fig. 6. X-ray photoemission spectroscopy spectrum of the SRO/STO thin film. The signals from the constituent elements are observed.

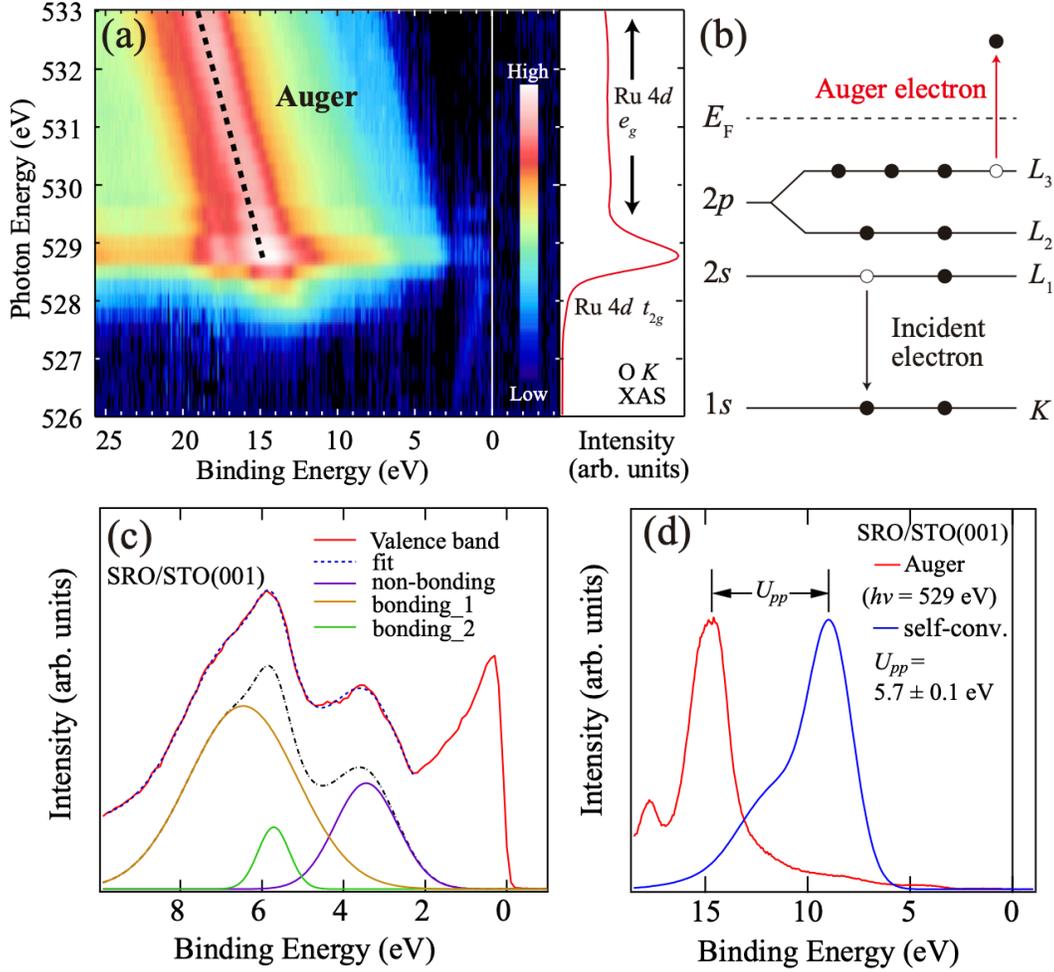

Fig. 7. Estimation of the ligand Coulomb interaction $U_{pp}$. (a) O $K$ RPES Mapping and O $K$ edge XAS spectrum. The dotted black line indicates the peak position of the Auger emission. (b) Schematic diagram of a *KLL* Auger process. (c) Decomposition of the valence-band PES spectrum. The electronic structure from 2.3 to 10 eV is decomposed into 3 Gaussians ("non-bonding", "bonding_1" and "bonding_2"). (d) Comparison between the Auger emission and the numerical self-convolution of the O 2$p$ states including the non-bonding and bonding (indicated with a purple and orange lines in (b)). Here, the contribution of the Ru 4$d$ state to the intensity of the bonding state has been extracted before the self-convolution. Black arrows indicate the difference between the peaks in the two spectra, which equals to the estimation of $U_{pp}$.